\documentclass[final,3p,times]{elsarticle}
\usepackage{amsmath}
\usepackage{amssymb}
\usepackage{lineno,hyperref}
\usepackage{natbib}

\biboptions{sort&compress}
\journal{Annals of Physics}
\hypersetup{colorlinks=true,linkcolor=blue,citecolor=blue,filecolor=blue,urlcolor=blue,breaklinks=true}
\RequirePackage{color}

\begin{document}

\begin{frontmatter}

\title{Quantum dynamics of relativistic bosons through nonminimal vector
square potentials}

\author[mymainaddress]{Luiz P. de Oliveira\corref{mycorrespondingauthor}}
\cortext[mycorrespondingauthor]{Corresponding author}
\ead{luizp@if.usp.br or oliveira.phys@gmail.com}

\address[mymainaddress]{Instituto de F\'{\i}sica, Universidade de S\~{a}o Paulo (USP),
\\05508-900, S\~{a}o Paulo, SP, Brazil}

\begin{abstract}
The dynamics of relativistic bosons (scalar and vectorial) through nonminimal vector square
(well and barrier) potentials is studied in the Duffin-Kemmer-Petiau (DKP)
formalism. We show that the problem can be mapped in effective Schr\"{o}%
dinger equations for a component of the DKP spinor. An oscillatory
transmission coefficient is found and there is total reflection.
Additionally, the energy spectrum of bound states is obtained and reveals the
Schiff-Snyder-Weinberg effect, for specific conditions the potential lodges
bound states of particles and antiparticles.
\end{abstract}


\end{frontmatter}

\section{Introduction}

The pioneering works of Duffin \cite{duffin}, Kemmer \cite{kem0, kem1} and
Petiau \cite{petiau} (DKP) gave rise to a rich formalism, similar to Dirac
theory, able to describe interactions of spin 0 and spin 1 bosons. Various
additional couplings, impossible to be explored in conventional Klein-Gordon
and Proca equations, gave rise to a large area of physical applications such
as describing the scattering of mesons by nuclei \cite{cla1, barret1, cla2},
the dynamics of bosons in curved space-time \cite{benito1} and noninertial
effect of rotating frames \cite{novo}, thermodynamic properties of bosons in
noncommutative plane \cite{novo2}, all works involving spin 0 systems.
Vector bosons in the expanding universe \cite{sucu} and in an Aharonov-Bohm
potential \cite{ahara} are examples of applications to spin 1 systems. The
Bose-Einstein condensate \cite{cas, abreu}, very special relativity \cite%
{hoff}, among others works, are applications for both spin systems.

The interest of one-dimensional potentials in DKP formalism has increased
significantly in recent decades, because the simplicity of equations
obtained provides great support for studying physical systems in higher
dimensions. Among the potentials used, we can highlight the double-step
potential \cite{eu1, eu2}, the DKP oscillator \cite{pot2, pot3}, the
inversely linear background \cite{pot4}, the mixed minimal-nonminimal vector
cusp potential \cite{pot5}.

In this spirit, the purpose of this article is to address the problem of
scalar and vector bosons subjected to a nonminimal vector square (well and
barrier) potential in the DKP formalism. We obtain a transmission
coefficient that shows oscillatory behavior, where we can observe the
resonance tunneling. Additionally, we obtain the energy spectrum of bound
states by a simple and transparent way. We show that the eigenenergies
obtained have great similarity to the problem of fermions in the same
potential, already explored in the literature \cite{luiz}.

\newpage

\section{A review of DKP equation}

The DKP equation for a free boson is given by \cite{kem0} (in natural units,
$\hbar =c=1$)%
\begin{equation}
\left( i\beta ^{\mu }\partial _{\mu }-m\right) \psi =0  \label{dkp}
\end{equation}%
\noindent where the matrices $\beta ^{\mu }$\ satisfy the algebra $\beta
^{\mu }\beta ^{\nu }\beta ^{\lambda }+\beta ^{\lambda }\beta ^{\nu }\beta
^{\mu }=g^{\mu \nu }\beta ^{\lambda }+g^{\lambda \nu }\beta ^{\mu }$
\noindent and the metric tensor is $g^{\mu \nu }=\,$diag$\,(1,-1,-1,-1)$.
The conserved four-current is given by $J^{\mu }=\bar{\psi}\beta ^{\mu }\psi
$\noindent $/2$ where the adjoint spinor $\bar{\psi}$ is given by $\bar{\psi}%
=\psi ^{\dagger }\eta ^{0}$ with $\eta ^{0}=2\beta ^{0}\beta ^{0}-1$. The
correct use of nonminimal interactions in the DKP equation can be found in
\cite{lui2}, where the continuity equation implies in conserved physical
quantities.

With nonminimal vector interactions, the DKP equation can be written as \cite%
{ume},
\begin{equation}
\left( i\beta ^{\mu }\partial _{\mu }-m-i[P,\beta ^{\mu }]A_{\mu }\right)
\psi =0  \label{dkp2}
\end{equation}%
where $P$ is a projection operator ($P^{2}=P$ and $P^{\dagger }=P$) in such
a way that $\bar{\psi}[P,\beta ^{\mu }]\psi $ behaves like a vector under a
Lorentz transformation as $\bar{\psi}\beta ^{\mu }\psi $ does. If the
potential is time-independent one can write $\psi (\vec{r},t)=\phi (\vec{r}%
)\exp (-iEt)$, where $E$ is the energy of the boson, the DKP equation becomes%
\begin{equation}
\left[ \beta ^{0}E+i\beta ^{i}\partial _{i}-\left( m+i[P,\beta ^{\mu
}]A_{\mu }\right) \right] \phi =0  \label{DKP10}
\end{equation}

\subsection{Scalar sector}

For the scalar bosons, we use the representation for the $\beta ^{\mu }$\
matrices given by \cite{ned1}%
\begin{equation}
\beta ^{0}=%
\begin{pmatrix}
\theta & \overline{0} \\
\overline{0}^{T} & \mathbf{0}%
\end{pmatrix}%
,\quad \beta ^{i}=%
\begin{pmatrix}
\widetilde{0} & \rho _{i} \\
-\rho _{i}^{T} & \mathbf{0}%
\end{pmatrix}%
,\quad i=1,2,3  \label{rep}
\end{equation}%
\noindent where%
\begin{eqnarray}
\ \theta &=&%
\begin{pmatrix}
0 & 1 \\
1 & 0%
\end{pmatrix}%
,\quad \rho _{1}=%
\begin{pmatrix}
-1 & 0 & 0 \\
0 & 0 & 0%
\end{pmatrix}
\notag \\
&&  \label{rep2} \\
\rho _{2} &=&%
\begin{pmatrix}
0 & -1 & 0 \\
0 & 0 & 0%
\end{pmatrix}%
,\quad \rho _{3}=%
\begin{pmatrix}
0 & 0 & -1 \\
0 & 0 & 0%
\end{pmatrix}
\notag
\end{eqnarray}%
\noindent $\overline{0}$, $\widetilde{0}$ and $\mathbf{0}$ are 2$\times $3, 2%
$\times $2 and 3$\times $3 zero matrices, respectively, while the
superscript T designates matrix transposition. Here the projection operator
can be written as \cite{gue} $P=\left( \beta ^{\mu }\beta _{\mu }-1\right)
/3=\mathrm{diag}\,(1,0,0,0,0)$. In this case $P$ picks out the first
component of the DKP spinor. The five-component spinor can be written as $%
\psi ^{T}=\left( \phi _{1},...,\phi _{5}\right) $ in such a way that the
time-independent DKP equation for a boson constrained moves along the $X$%
-axis, restricting ourselves to potentials depending only on $x$, decomposes
into
\begin{equation}
\left( \frac{d^{2}}{dx^{2}}+E^{2}-m^{2}+A_{0}^{2}-A_{1}^{2}+\frac{dA_{1}}{dx}%
\right) \phi _{1}=0  \label{ulla}
\end{equation}%
\begin{equation}
\phi _{2}=\frac{1}{m}\left( E+iA_{0}\right) \,\phi _{1}  \label{dkp4}
\end{equation}%
\begin{equation}
\phi _{3}=\frac{i}{m}\left( \frac{d}{dx}+A_{1}\right) \phi _{1},\quad \phi
_{4}=\phi _{5}=0
\end{equation}%
And the conserved currents have the form%
\begin{equation}
J^{0}=\frac{E}{m}\,|\phi _{1}|^{2},\quad J^{1}=\frac{1}{m}\text{Im}\left(
\phi _{1}^{\ast }\,\frac{d\phi _{1}}{dx}\right) .  \label{corrente4}
\end{equation}

\subsection{Vector sector}

Using the representation of $\beta ^{\mu }$\ matrices for vector bosons,
given by \cite{ned2}:

\begin{equation}
\beta ^{0}=%
\begin{pmatrix}
0 & \overline{0} & \overline{0} & \overline{0} \\
\overline{0}^{T} & \mathbf{0} & \mathbf{I} & \mathbf{0} \\
\overline{0}^{T} & \mathbf{I} & \mathbf{0} & \mathbf{0} \\
\overline{0}^{T} & \mathbf{0} & \mathbf{0} & \mathbf{0}%
\end{pmatrix}%
,\quad \beta ^{i}=%
\begin{pmatrix}
0 & \overline{0} & e_{i} & \overline{0} \\
\overline{0}^{T} & \mathbf{0} & \mathbf{0} & -is_{i} \\
-e_{i}^{T} & \mathbf{0} & \mathbf{0} & \mathbf{0} \\
\overline{0}^{T} & -is_{i} & \mathbf{0} & \mathbf{0}%
\end{pmatrix}%
\end{equation}%
\noindent where $s_{i}$ are the 3$\times $3 spin-1 matrices, $e_{i}$ are the
1$\times $3 matrices $\left( e_{i}\right) _{1j}=\delta _{ij}$ and $\overline{%
0}=%
\begin{pmatrix}
0 & 0 & 0%
\end{pmatrix}%
$, while $\mathbf{I}$ and $\mathbf{0}$ designate the 3$\times $3 unit and
zero matrices, respectively, the time-independent DKP equation (see Ref.
\cite{eu2}) can be written in the simpler form
\begin{equation*}
\left( \frac{d^{2}}{dx^{2}}+k_{\sigma }^{2}\right) \phi _{I}^{\left( \sigma
\right) }=0
\end{equation*}%
\begin{equation}
\phi _{II}^{\left( \sigma \right) }=\frac{E+i\sigma A_{0}}{m}\,\phi
_{I}^{\left( \sigma \right) }  \label{poup}
\end{equation}%
\begin{equation*}
\phi _{III}^{\left( \sigma \right) }=\frac{i}{m}\left( \frac{d}{dx}+\sigma
A_{1}\right) \phi _{I}^{\left( \sigma \right) },\quad \phi _{8}=0
\end{equation*}%
where%
\begin{equation}
k_{\sigma }^{2}=E^{2}-m^{2}+A_{0}^{2}-A_{1}^{2}+\sigma \frac{dA_{1}}{dx}
\label{haah}
\end{equation}%
and $\sigma $ is the polarization of vector boson states, i.e, $\sigma =-$
for transverse and $\sigma =+$ for longitudinal. In this representation, the
projector is give by $P=\,\beta ^{\mu }\beta _{\mu }-2=\mathrm{diag}%
\,(1,1,1,1,0,0,0,0,0,0)$ \cite{vij}. Now the components of the four-current
are%
\begin{equation}
J^{0}=\frac{E}{m}\sum\limits_{\sigma }|\phi _{I}^{\left( \sigma \right)
}|^{2},\quad J^{1}=\frac{1}{m}\,\text{Im}\sum\limits_{\sigma }\phi
_{I}^{\left( \sigma \right) \dagger }\,\frac{d\phi _{I}^{\left( \sigma
\right) }}{dx}.  \label{CUR2}
\end{equation}

We can see, from ($11$) and ($12$), that the solution for vector
sector consists in searching solutions for two Klein-Gordon-like equations
for $\phi _{I}^{\left( \sigma \right) }$, but $\phi _{I}^{\left( +\right) }$%
and $\phi _{I}^{\left( -\right) }$ are not independents because $E$ is the
constraint that appears in both equations. Cardoso and collaborators \cite%
{jpa} had already alerted that the solutions for the spin 1 sector of the
DKP equation, if they really exist, can be obtained from a restrict class of
solutions of the spin 0 sector. There is not surprise because in the absence
of any interaction, the free Proca fields obey a free Klein-Gordon equation
with a constraint on the components of the Proca field.

\section{The nonminimal vector square potentials}

The square (well and barrier) potentials are given by

\begin{eqnarray}
A_{\mu }(x) &=&b_{\mu }V_{0}g(x)\text{ \ with \ }\mu =0,1, \\
g(x) &=&\frac{1}{2}\left[ sgn(x-a)-sgn(x+a)\right]
\end{eqnarray}%
where $b_{\mu }V_{0}$ is a positive (negative) constant for wells (barriers)
with energy dimension and $sgn(x)$ is the sign function.

\subsection{Scalar bosons}

With this potential, eq. (\ref{ulla}) becomes%
\begin{equation}
\frac{d^{2}\phi _{1}}{dx^{2}}+\left\{ E^{2}-m^{2}+jV_{0}^{2}g(x)+\frac{%
b_{1}V_{0}}{2}\left[ \delta \left( x-a\right) -\delta \left( x+a\right) %
\right] \right\} \phi _{1}=0  \label{eq0}
\end{equation}%
where $\delta \left( x\right) =d\theta \left( x\right) /dx$ is the Dirac
delta function and $j\equiv b_{1}^{2}-b_{0}^{2}$. We turn our attention to
scattering states so that the solutions describing spinless bosons coming
from the left can be written as
\begin{equation}
\phi _{1}(x)=\left\{
\begin{array}{cc}
Ae^{+i\xi \frac{x}{a}}+Be^{-i\xi \frac{x}{a}} & \text{\textrm{for }}x<-a \\
&  \\
Ce^{+i\eta \frac{x}{a}}+De^{-i\eta \frac{x}{a}} & \text{\textrm{for }}|x|<a
\\
&  \\
Fe^{+i\xi \frac{x}{a}} & \text{\textrm{for }}x>a%
\end{array}%
\right.  \label{eq50}
\end{equation}%
where%
\begin{equation}
\xi =a\sqrt{E^{2}-m^{2}},\quad \eta =\sqrt{\xi ^{2}-j\mathcal{\upsilon }^{2}}%
,\quad \mathcal{\upsilon }=aV_{0}
\end{equation}

The group velocity of the waves described above is given by

\begin{equation}
v_{g}=\frac{dE}{d(\xi/a)}=\pm \frac{\xi/a}{\sqrt{(\xi/a)^{2}+m^{2}}}
\end{equation}%
where the double signal is related to boson propagation direction.

Then, $\phi _{1}$ describes an incident wave moving to the right ($\xi $ is
a real number) and a reflected wave moving to the left with%
\begin{equation}
J^{1}\left( x<-a\right) =\frac{\xi }{am}\left( |A|^{2}-|B|^{2}\right)
\end{equation}%
and a transmitted wave moving to the right with%
\begin{equation}
J^{1}\left( x>a\right) =\frac{\xi }{am}|F|^{2}
\end{equation}

We demand $\phi _{1}$, to be continuous at $x=\pm a$, i. e.%
\begin{equation}
\lim_{\varepsilon \rightarrow 0}\left. \phi _{1}\right\vert _{x=\pm
a-\varepsilon }^{x=\pm a+\varepsilon }=0  \label{cont}
\end{equation}%
and the connection formula between $d\phi _{1}/dx$ at the right and $d\phi
_{1}/dx$ at the left can be summarized as%
\begin{equation}
\lim_{\varepsilon \rightarrow 0}\left. \frac{d\phi _{1}}{dx}\right\vert
_{x=\pm a-\varepsilon }^{x=\pm a+\varepsilon }=\mp \frac{b_{1}\mathcal{%
\upsilon }}{2a}\,\phi _{1}(\pm a).  \label{discont}
\end{equation}

Omitting the algebraic details, we obtain the following amplitudes%
\begin{eqnarray}
r &=&\frac{e^{-2i\xi }\left[ if(\xi )-\gamma (\xi )\right] }{\cos (2\eta
)-if(\xi )\sin (2\eta )}\sin (2\eta )  \notag \\
&&  \label{amp} \\
t &=&\frac{e^{-2i\xi }}{\cos (2\eta )-if(\xi )\sin (2\eta )}.  \notag
\end{eqnarray}%
where we have defined%
\begin{eqnarray}
\gamma (\xi ) &\equiv &\frac{vb_{1}+2i\xi }{2\eta }, \\
f(\xi ) &\equiv &\frac{4\left( \xi ^{2}+\eta ^{2}\right) +b_{1}^{2}v^{2}}{%
8\eta \xi }.
\end{eqnarray}

In order to determinate the reflection and transmission coefficients we use
the charge current densities $J^{1}\left( x<-a\right) $ and $J^{1}\left(
x>a\right) $. The $x$-independent current density allow us to define the
reflection and transmission coefficients as%
\begin{equation}
R=\left\vert r\right\vert ^{2},\quad T=\left\vert t\right\vert ^{2}
\end{equation}%
with $R+T=1$. Therefore,
\begin{eqnarray}
R &=& \frac{|f(\xi)|^{2}+ 2i\mbox{Im}|\gamma(\xi)f^{*}(\xi)|+|\gamma(\xi)|^{2}}
{1+\left[f(\xi)^{2} -1\right ]\sin^{2}(2\eta)-i\sin(4\eta)\mbox{Im}
\left[ f(\xi) \right ]} \sin^{2}(2\eta) \, , \nonumber \\
\left. \right.  \\
T &=& \frac{1}{1+\left[f(\xi)^{2} -1\right ]\sin^{2}(2\eta)-i\sin(4\eta)\mbox{Im}
\left[ f(\xi) \right ]}  \nonumber
\end{eqnarray}%
The figure 1 shows the profiles of reflection and transmission coefficients
for $v=2,$ $j=b_{1}=1$. As expected, $T$ does not depend of sign of $b_{1}$.
Notice that $T\rightarrow 1$ when

\begin{equation}
\eta =\frac{(N+1)\pi }{2}\text{, \ \ with \ \ }N=0,1,2,\ldots
\end{equation}%
and that there is a resonance transmission ($T=1$) whenever

\begin{equation}
\left\vert \xi \right\vert =\sqrt{\frac{(N+1)^{2}\pi ^{2}}{4}+jv^{2}}.
\end{equation}

Another expected result is $T\rightarrow 1$ when $E\rightarrow \pm \infty $,
observed in the figure 1 (logically, there is a symmetry $E\rightarrow -E$ ).

\begin{figure}[!ht]
\centering
\includegraphics[scale=0.5]{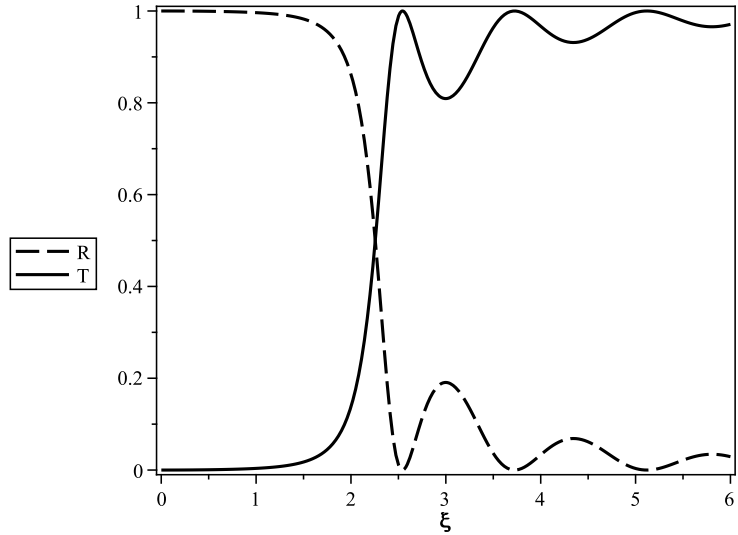}
\caption{Reflection and transmission coefficients for $v=2$ and $m=1$.}
\label{figure 1}
\end{figure}

We can observe a great similarity between $T$ for fermions \cite{luiz} and
bosons in the same potential, i.e, both do not depend on delta function
localization and have the same resonance points. This is well understood
because both have the same effective Schr\"{o}dinger equations for spinor
components in square potentials. However, there is no total reflection for
fermions.

Additionally, we can obtain the bound state solutions with the prescription $%
\xi \rightarrow i\left\vert \xi \right\vert $ in the transmission amplitude.
In this clear way, the bound state spectrum is obtained from poles of $t$,
and the wave functions have the same form as (13) with $A=0$. As expected,
there are bound state solutions just for $j<0$ and $b_{1}=1$, because the
effective equation show a attractive potential between the two delta
function potentials. It is like fermions bind by the same effective
potential \cite{luiz}.

Therefore, for $j=-1$,
\begin{equation}
\frac{8\sqrt{v^{2}-\left\vert \xi \right\vert ^{2}}}{5v^{2}-8\left\vert \xi
\right\vert ^{2}}\left\vert \xi \right\vert =\tan (2\sqrt{v^{2}-\left\vert
\xi \right\vert ^{2}})
\end{equation}%
is the quantization condition for the problem of bound scalar bosons by an
effective well potential with attractive and repulsive deltas at the
borders. The condition (27) can be solved by the graphical method and the
numerical solutions are show in the figure 2. The nonminimal vector coupling
is impossible in Klein-Gordon equation and we can observe this fact in the
bound states spectrum. In the Ref. \cite{tat}, the authors solve the
minimally coupling Klein-Gordon equation with the same square potential, but
the spectrum is completely different from the one obtained in this work.
This reveals the great applicability of DKP formalism for describing
physical systems with many coupling possibilities. The presence of delta
functions at the borders has impact under the parity of the solutions, i.e,
there is a symmetry breaking due to the delta potentials. The effective
potential has no defined parity which implies in solutions without defined
parity.

From the figure 2 we can see that the energy levels decay more rapidilly
with the $v$ increasing. This can be explained by the limit $v\rightarrow
\infty $, which provides that there are more bound states with the increase
of depth well. However, in the limite case

\begin{equation}
V_{0}\rightarrow \infty \text{ \ and \ }a\rightarrow 0
\end{equation}%
we obtain just an atractive delta potential at the origin $x=0,$ which
provides only one bound level \cite{grif}. In figure 3, we have the
behaviour of $E$ in function of the lenght $a$ for $V_{0}/m=50$ (strong
potential). In the limit $a\rightarrow 0,$ we can see the expected only one
bound state level.

From figures 2 and 3 we can conclude that there is no Klein's paradox for
this configuration, i.e, particle levels penetrating in the antiparticle
\emph{continuum} region with the increasing of $V_{0}$ and $a$.

An interesting relativistic phenomena observed from bound state spectrum is
the Schiff-Snyder-Weinberg (SSW) effect \cite{ssw}. This effect occurs when
an attractive well for particles in a critical depth lodges bound states of
particles and antiparticles in the Klein-Gordon equation. Popov \cite{popov}
suggests that the effect is characteristic of short-range and depth
potentials. Our nonminimal vector potential, in the limits given by (28),
contain all the characteristics to exhibit the SSW effect in the DKP
formalism. From figures 2 and 3, we can see this phenomena because the
spectrum is symmetrical with respect to $E$. We know that the minimally
coupled case, the DKP equation reveals the SSW effect only with intense
vector potential. However, in the nonminimal vector case, the SSW effect
occurs independent of intensity potential. The explanation for this
difference that makes our results be expected is that the DKP equation, in
the presence of nonminimal vector interactions, is invariant under charge
conjugation. Therefore, the square potential lodges bound states of
particles and antiparticles, independent of intensity potential, as seen in
figures 2 and 3.

\subsection{Vector bosons}

Vector bosons are subject to the effective potential

\begin{equation}
V_{eff}^{\sigma }(x)=\frac{1}{2m}\left( A_{1}^{2}-A_{0}^{2}-\sigma \frac{%
dA_{1}}{dx}\right) ,
\end{equation}%
which there is $\sigma -$dependence. However, for our problem, we can see
that the polarization is only relate to the localization of delta functions
at the borders of the square potential, i.e, if the delta potentials are
attractive or repulsive. The results for scalar bosons does not depend on
the position of delta functions. Therefore, the same results ($R$, $T$ and
bound states spectrum) of scalar bosons are obtained for vector bosons
subjected to the same square potential, for both possibilities $\sigma =\pm
. $ We must remember that these results were expected since solutions for
spin $1$ bosons can be obtained from spin $0$ bosons as pointed in \cite{jpa}%
.

The square potentials are simple models much explored in quantum physics
books, for example \cite{grif}. Among various applications, L. Schiff \cite%
{schiff} used a square well potential in the Klein-Gordon equation to bind
the di-pions, providing a satisfactory account of the observed $P$-wave
pion-pion scattering. In the Ref. \cite{dahm}, the authors used a square
well potential as a model that allows calculating the phonon spectral
function analytically in the nuclear magnetic resonance (NMR) relaxation.
The authors were able to estimate an absolute value for the expected peak
position of the NMR relaxation rate near the experimental data. An
application of barrier potentials can be found in Ref. \cite{bennett}, where
the authors studied the tunneling spectroscopy of collective excitations.
Therefore, our results can be applied to all spinless bosons systems and the
many couplings in the DKP formalism enable this.

\begin{figure}[tbh]
\centering
\includegraphics[scale=0.4]{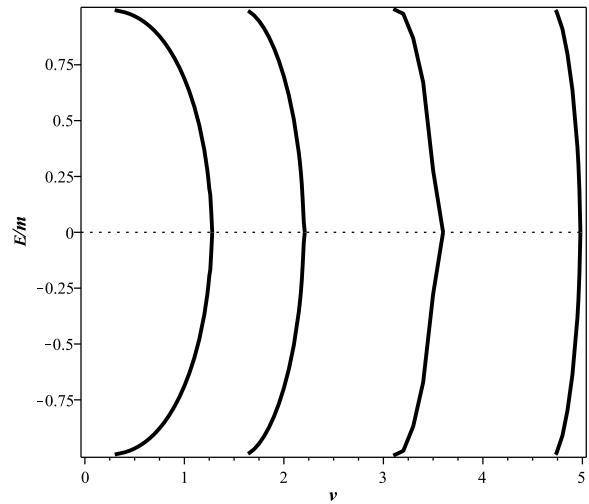}
\caption{Energy levels of bound states.}
\label{figure 2}
\end{figure}

\begin{figure}[!htb]
\centering
\includegraphics[scale=0.4]{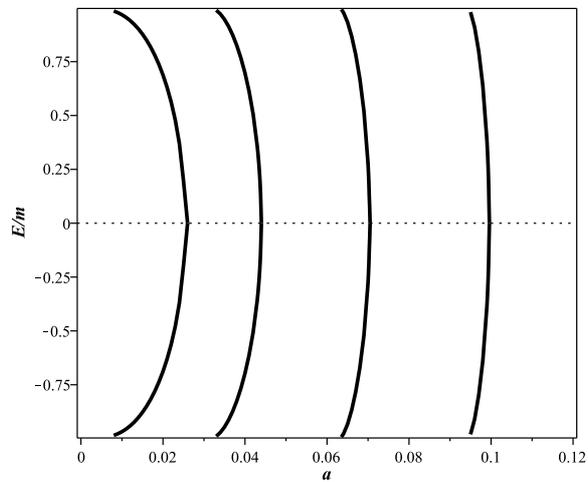}
\caption{Energy levels in function of $a$ for $V_{0}=50$.}
\label{figure 3}
\end{figure}

\newpage

\section{Conclusions}

A physical system containing spinless bosons subject to the nonminimal
vector square potentials is studied in the DKP formalism. The scattering
states reveal an oscillatory transmission coefficient and the interesting
particle problem embedded in a delta function is obtained. We can observe
that there are bound state solutions only when the time component of square
potential is more intense than its spatial component $(j<0)$. The energy
bound states were obtained in a simple way from the poles of transmission
amplitude. The parity of bound solutions is broken, similar to the fermions
problem in the same potential which was already discussed in the literature
\cite{luiz}. The bound states spectrum reveals the Schiff-Snyder-Weinberg
effect \cite{ssw}, confirming the Popov's work \cite{popov} and its
applicability to the DKP formalism. From a simple analysis, we can obtain the
solutions for vector bosons from scalar bosons results.

\section*{Acknowledgments}

This work was supported by means of funds provided by Coordena\c{c}\~{a}o de
Aperfei\c{c}oamento de Pessoal de N\'{\i}vel Superior (CAPES). The author is
grateful to the anonymous referee for the suggestions of this work. The
author would like to thank professor L.B. Castro for useful discussions.


\begin{thebibliography}{99}
\bibitem{duffin} R. J. Duffin, Phys. Rev. \textbf{54}, 1114 (1938).

\bibitem{kem0} N. Kemmer, Proc. R. Soc. A \textbf{166}, 127 (1938).

\bibitem{kem1} N. Kemmer, Proc. R. Soc. A \textbf{173}, 91 (1939).

\bibitem{petiau} G. Petiau, Acad. R. Belg., A. Sci. M\'{e}m. Collect.
\textbf{16}, 2 (1936).

\bibitem{cla1} B. C. Clark \textit{et al}., Phys. Rev. Lett. \textbf{55},
592 (1985).

\bibitem{barret1} R. C. Barret and Y. Nedjadi, Nucl. Phys. A \textbf{585},
311c (1995).

\bibitem{cla2} B. C. Clark \textit{et al}., Phys. Lett. B \textbf{427}, 231
(1998).

\bibitem{benito1} L. B. Castro, Eur. Phys. J. C \textbf{75}, 287 (2015).

\bibitem{novo} L. B. Castro, Eur. Phys. J. C \textbf{76}, 61 (2016).

\bibitem{novo2} Z. Wang, Z. Long, C. Longand, W. Zhang, Advances in High
Energy Physics \textbf{2015}, Article ID 901675, (2015).
http://dx.doi.org/10.1155/2015/901675

\bibitem{sucu} Y. Sucu and N. Unal, Eur. Phys. J. C \textbf{44}, 287 (2005).

\bibitem{ahara} L. B. Castro and E. O. Silva, arXiv:1507.07790 (2015).

\bibitem{cas} R. Casana, V. Y. Fainberg, B. M. Pimentel, J. S. Valverde,
Phys. Lett. A \textbf{316}, 33 (2003).

\bibitem{abreu} L. M. Abreu, A. L. Gadelha, B. M. Pimentel, E. S. Santos,
Physica A: Statistical Mechanics and its Applications \textbf{419}, 612
(2015).

\bibitem{hoff} R. M. T. Cavalcanti, J. M. Hoff da Silva, R. A. da Rocha,
Eur. Phys. J. Plus \textbf{129}, 246 (2014).

\bibitem{eu1} L. P. de Oliveira and A. S. de Castro, Can. J. Phys. \textbf{90%
}, 481 (2012).

\bibitem{eu2} L. P. de Oliveira and A. S. de Castro, Int. J. Mod. Phys. E
\textbf{24}, 1550031 (2015).

\bibitem{pot2} D. A. Kulikov, R. S. Tutik, and A. P. Yaroshenko, Mod. Phys.
Lett. A \textbf{20}, 43 (2005).

\bibitem{pot3} L. B. Castro and A. S. de Castro, Phys. Lett. A \textbf{375},
2596 (2011).

\bibitem{pot4} A. S. de Castro, J. Math. Phys. \textbf{51}, 102302 (2010).

\bibitem{pot5} A. S. de Castro, J. Phys. A \textbf{44}, 035201 (2011).

\bibitem{luiz} L. P. de Oliveira and L. B. Castro, Ann. Phys. \textbf{364},
99 (2016).

\bibitem{lui2} L. B. Castro and L. P. de Oliveira, Advances in High Energy
Physics \textbf{2014}, Article ID 784072, (2014).
http://dx.doi.org/10.1155/2014/784072

\bibitem{ume} H. Umezawa, \textit{Quantum Field Theory}. North-Holland,
Amsterdam (1956).

\bibitem{ned1} Y. Nedjadi and R. C. Barret, J. Phys. G \textbf{19}, 87
(1993).

\bibitem{gue} R. F. Guertin and T. L. Wilson, Phys. Rev. D \textbf{15}, 1518
(1977).

\bibitem{ned2} Y. Nedjadi and R. C. Barret, J. Math. Phys. \textbf{35}, 4517
(1994).

\bibitem{vij} B. Vijayalakshmi, M. Seetharaman, and P. M. Mathews, \textit{%
J. Phys. A} \textbf{12, }665 (1979).

\bibitem{jpa} T. R. Cardoso, L. B. Castro and A. S. de Castro, J. Phys. A
\textbf{43}, 055306 (2010).

\bibitem{tat} T. R. Cardoso and A. S. de Castro, Rev. Bras. Ens. F\'{\i}s.
\textbf{30}, 2606 (2008).

\bibitem{grif} D. J. Griffiths, \textit{Introduction to Quantum Physics},
Pearson Prentice Hall, Upper Saddle River (2004).

\bibitem{ssw} L.I. Schiff, H. Snyder and J. Weinberg, Phys. Rev. \textbf{57}%
, 315 (1940).

\bibitem{popov} V.S. Popov, Sov. Phys. JETP \textbf{32}, 526 (1971).

\bibitem{schiff} L. I. Schiff, Phys. Rev. \textbf{125}, 777 (1962).

\bibitem{dahm} T. Dahm and K. Ueda, J. Phys. Chem. Solids \textbf{69}, 3160
(2008).

\bibitem{bennett} A. J. Bennett, C. B. Duke and S. D. Silvertein, Phys. Rev
\textbf{176}, 969 (1968).
\end{thebibliography}
\end{document}